# Evaluating User Perception of Multi-Factor Authentication A Systematic Review

S. Das, B. Wang, Z. Tingle, and L. Jean Camp
School of Informatics, Computing, and Engineering
Indiana University Bloomington
{sancdas, bw10, zatingle, ljcamp} @iu.edu

## Abstract

Security vulnerabilities of traditional single factor authentication has become a major concern for security practitioners and researchers. To mitigate single point failures, new and technologically advanced Multi-Factor Authentication (MFA) tools have been developed as security solutions. However, the usability and adoption of such tools have raised concerns. An obvious solution can be viewed as conducting user studies to create more user-friendly MFA tools. To learn more, we performed a systematic literature review of recently published academic papers ($N = 623$) that primarily focused on MFA technologies. While majority of these papers ($m = 300$) proposed new MFA tools, only 9.1% of papers performed any user evaluation research. Our meta-analysis of user focused studies ($n = 57$) showed that researchers found lower adoption rate to be inevitable for MFAs, while avoidance was pervasive among mandatory use. Furthermore, we noted several reporting and methodological discrepancies in the user focused studies. We identified trends in participant recruitment that is indicative of demographic biases.

## Keywords

Authentication, Multi-Factor Authentication, Passwords, Systematic Literature Review, User Studies, User Experience, User Evaluation, Human-Factors.

## 1. Introduction

Online user presence increased considerably in the last decade (Kemp 2017), where in 2018, 89% adults in the U.S. reported using internet daily (*Statistic* 2018). Such exponential growth in users and data (Patil & Seshadri 2014) has warranted security practitioners to become more concerned with online data security (Al Hasib 2009) and access control issues (Cuzzocrea 2014). Traditional single-factor authentication (SFA), such as textual passwords (O'Gorman 2003) or a Personal Identification Number (PIN) (Dodge & Kitchin 2005), are intended for user identity verification (Hinton & Vandenwauver 2009). However, risk assessments of SFA have disclosed several vulnerabilities to security attacks, such as, brute force (Owens & Matthews 2008), dictionary (Sood et al. 2009), malware (Fovino et al. 2009), Keyloggers (Kim & Hong 2011), and others (Tari et al. 2006). As a solution, Multi-Factor Authentication (MFA) creates multiple layer of security in addition to the single sign-ons (Chaudhari et al. 2011).

Irrespective of increased data security (Labana et al. 2013), MFA tools have several usability challenges (De Cristofaro et al. 2013), such as a user's lack of motivation (*Das et al. 2019*.), risk trade-off understanding (Tari et al. 2006), and presence of non-intuitive user interfaces (Braz & Robert 2006). Conducting user studies (Keith et al. 2007) to provide proper risk alignment have been proven to be effective in improving digital security through adoption. For instance, Das et al., following a think-aloud protocol, studied user behavior of two-factor authentication (2FA) and provided actionable recommendations which enhanced usage experience and in turn adoption of 2FA (Das et al. 2018*b*). Studies on the usability of authentication methods is often undervalued by security practitioners (Egelman et al. 2014). Thus, a detailed systematic literature review is imperative to understand where we can improve as a research community (Das, Kim, Tingle & Camp 2019). To our surprise, our research revealed that only 9.1% of our collected studies which focused on MFA, conducted any user evaluation. The aim of our study is to improve user adoption of MFA and how we can utilize the pre-existing research to improve future study designs.

For our research, began by performing a systematic literature review partially adapted from the work of Stowell et al. (Stowell et al. 2018). We then compiled a set of recently studied academic papers containing keywords such as, multi-factor authentication, two factor authentication, and password. Using these keywords, we derived our set of literature works from four different academic databases: Google Scholar, ACM, Science Direct, and IEEE. We then derived sub-lists from these papers to obtain a sample of papers focusing on user studies for meta-analysis. Our findings show that in addition to the lack of user evaluation, there are several issues such as, lack of population diversity in these studies and exclusion of expertise knowledge on evaluation of usage statistics. We acknowledge that all of the studied papers were rich in their research contribution, however, our aim is to further improve the study designs for better future research practices.

## 2. Related Work

MFA involves multi-layer authentication scheme to mitigate risks of single factor sign-ons, such as, password breaches and unauthorized access of trusted devices (Hwang et al. 2002). Previous research on MFA primarily focused on the technological improvement of authentication and access control to address existing weakness in various areas such as security and compatibility with applications (Chayanam et al. 2012). However, the usability, adoption, and alignment with user risk perception remains a question (Das et al. 2018*a*). While new authentication methods have been found more interesting to explore, previous studies also have intensively evaluated existing MFA on the aspect of speed, simplicity (user actions) and authentication error rates on the user side (Wang & Wang 2018; Nag et al. 2014; Abo-Zahhad et al. 2016). However, usability of high touch and low tech schemes, still remains a challenge (Das et al. 2018*a*). Our study revealed several reporting issues which occur in current usability studies, which might generate inconclusive results.

An analysis of user studies provides the necessary information for improvement of a user's multi-factor authentication experience (Liou & Bhashyam 2010). Systematic literature review often helps in understanding the literature gap to pave future study directions (Brereton et al. 2007). Our systematic literature review is inspired by Stowell et al.'s work, "Designing and Evaluating mHealth Interventions for Vulnerable Populations: A Systematic Review". They begin their literature review by collecting a wide-range collection of papers related to mHealth (Kay et al. 2011) technology studies. Information such as demographics and types of studies conducted was gathered from each extracted paper. By recording these findings, they were able to understand the existing literature and pave the future scope of such research. Our study provides a survey of existing literature that identifies the current trends that user studies are going toward. In doing so, we aim to provide a foundation for more effective user studies in multi-factor authentication in the future.

## 3. Methods

We adapted the study methodology for the literature review of Multi-Factor Authentication from Stowell et al.'s work (Stowell et al. 2018). Additionally, we modified the protocol to better fit our research needs. Methods utilized in our research involve the following steps: (1) Data Collection through database search, (2) Data Screening involving: Title screening,

Abstract screening, Full-Text Screening, and (3) Data Extraction through Publish or Perish [1]. We started our data collection by generating a large sample of papers related to a set of keywords from four major databases: ACM, IEEE Xplore, Google Scholar and Science Direct. We also performed a Quality Assessment of the papers to ensure that they met our inclusion or exclusion criteria.

Papers were included if they met the following criteria: (1) The paper published in a peer-reviewed conferences or journals, (2) The primary language used to write the paper was English (3) The full text was available over Publish or Perish for us to performed detailed analysis. For the papers where we could not find in Publish or Perish we tried obtaining the full text for in the databases mentioned earlier by manually going through it. (4) User studies papers, as we need to perform detailed analysis on Human Subjects, this inclusion criteria were added during the meta-analysis. (5) Papers that primarily focused on authentication technologies. Such as password, 2FA, and MFA tool and technologies. (6) Papers published in 2018. We particularly focused on 2018 since we wanted to capture the user adoption and perception issues for the current technologies. This was also done to funnel our research for detailed insights of user focused studies.

We also followed the exclusion criteria for quality assessment. Papers were excluded if: (1) The full text was not available as of December 2018. (2) Presented as semi-finished work, such as posters, extended abstracts, or work in progress papers. A meta-analysis was conducted to determine if article contained specific demographic information (Gender, Age, etc.) or research information such as security background, survey, interview or experiment. For our meta-analysis, we excluded papers if: (1) Had insufficient details of research intention through their recruitment procedure. (2) Did not study the user behavior through any form of usability evaluation as we performed thematic analysis and segregated our collected papers into user and non-user focused studies. Our procedure consisted of applying the following filters sequentially via the "Advanced Search" feature of each database and an overall description of our data collection, screening, quality assessment, and analysis can be sketched in Figure 1.

## 4. Findings

During our systematic literature review, we investigated the existing set of literature based around user studies in multi-factor authentication for paving the path for future studies by underlining existing gaps in research. Below are major findings we've discovered during the research.

### 4.1. Overall Analysis

We conducted a thorough coding analysis of the 623 collected papers that revealed trends throughout. We then categorized these codes under six primary categories consistent with a specified theme. Table 1 gives the overall distribution of the studies. These codes were not mutually exclusive.

---

[1] https://harzing.com/resources/publish-or-perish

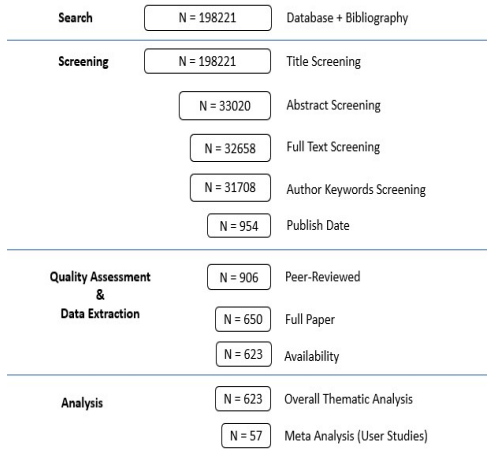

**Figure 1:** Overview of the Study Methodology and Design

| Overall Categories | | |
|---|---|---|
| Cyber Threat Testing | 246 | 39.5% |
| Traditional Authentication Schemes | 143 | 22.9% |
| Industry Manufacturers | 35 | 5.6% |
| New Authentication Technologies | 300 | 48.2% |
| User Based Studies | 57 | 9.1% |
| Organizational Implementation | 15 | 2.4% |

**Table 1:** Overall Categories of Collected Research Papers

Majority of our collected sample set ($N = 48.2\%$) focused primarily on proposing new MFA technologies. Across all these studies, Graphical user authentication was the common theme. This indicated the security trend is going forward towards interactive authentication schemes and in turn creating more user-friendly tech. Several companies, such as Duo [2], Yubico [3], Okta [4] and others focus on creating MFA technologies. We wanted to analyze if studies were focused on testing of evaluating the technological products from these manufacturers. Out of the 623 papers, only 50 of them discussed about tools that has already been developed. We found that Yubico was one of the most studied organization among MFA technology vendors (Reynolds et al. 2018; Das et al. 2018b), where in both studies, a two-phase user study was implemented, and recommendations enhanced the Yubico usability and adoption to a considerable amount. It will be interesting to delve further in future researches the application perspective of the MFA tools for larger industries. Most of the papers collected (39.5%), explored the threats involved with single sign factor authentication and how MFA can be implemented to solve those issues. Traditional authentication schemes such as passwords contribute to the SFA. Although this research was primarily focused on MFA, it is important for us to note the password analysis which the researchers focused on. The works focused primarily on SFA vulnerabilities and and focused on the mental models of users in password creation and management- whereby users were tested on how they both develop passwords and how they keep track/recall them. Account security is highly dependent on the creation of effective, secure passwords that are not uniformly used across multiple websites. Studies that were conducted to understand passwords and traditional authentication are more concerned with password recall, how users create passwords, and overall password strength. Only 2.4% of the research work focused on any organizational implementation of the MFA. Here we considered both Universities and Industrial organizations, however majority of the work only focused on university implication, despite being the industry as a major source of workforce and data repository. Industrial implication is often understudied, primarily because the data policies of the industry as well as lack of contribution from the organizations itself and the recruitment can be challenging. However, to provide an overall adoption strategy of MFA such studies are extremely critical.

---

[2] https://duo.com/

[3] https://www.yubico.com/

[4] https://www.okta.com/

## 4.2. Meta-Analysis

As mentioned earlier, usability and adoption is often challenging for MFA and performing an overall analysis is helpful to learn about the current research trends of MFA, however, we wanted to delve deeper to explore the user studies. We really appreciate the extensive work in this field, however, our analysis of ($n = 57$) papers, revealed several user study biases to inconsistent reporting issues. Our analysis focused on the participant pool, study design, execution strategies, and findings to pave future research directions.

### 4.2.1. Risk Perception Analysis

Risk perception analysis is extremely helpful in understanding the risk in security challenges. We identified majority researches on risk perception are focused on usability and password memorability. Table 2 shows the different types of risk analysis the studies performed; tool risk trade-off understanding was studied for 5% of the paper which was an interesting finding since many research claim that there is a misalignment of user risk perception with tool's utility. Nudging was considered as a primary method to interject into the risk mental models of the users.

| Risk Perception | |
|---|---|
| Cognitive Differences | 14 (24.6%) |
| Nudge Theory | 8 (14.0%) |
| Password Memorability | 15 (26.3%) |
| Tool Risk Trade-off | 15 (26.3%) |
| Security Motivation | 15 (26.3%) |
| Understanding Password Security | 25 (44.0%) |
| Usability Study | 16 (28.1%) |
| User Risk Models | 9 (15.8%) |

**Table 2:** Distribution of studies which observed User Risk Perception

### 4.2.2. Traditional Authentication Studies

While MFA is gradually gaining popularity, password authentication still dominates the area of single-factor authentication, as well as the first factor in MFA authentication. We saw that 16% of the user studies focused on understanding the password security understanding of the users. We found that security researchers are particularly interested in the password creation and management shown in the table 3.

| Traditional Single-factor Authentication | |
|---|---|
| Conventional Passwords | 8 (14.0%) |
| Password Creation | 12 (21.1%) |
| Password Management | 16 (28.1%) |
| Password Meter | 8 (14.0%) |
| Password Cracking | 2 (3.5%) |
| Password Guessability | 2 (3.5%) |
| Student Created Password | 3 (5.3%) |

**Table 3:** Distribution of studies which discussed password studies

### 4.2.3. Participant Recruitment Biases

Participant biases was a major concern while we performed our analysis. A majority of the user studies divulged throughout the course of this study gather their participants primarily through university settings (Naiakshina et al. 2018; Becker et al. 2018). These participants are often college aged, 18 to 22 years old, and by effect more technologically literate (Constantinides et al. 2018). Some of these studies even utilize computer science students and individuals who are employed professionally (Renaud & Zimmermann 2018). While convenient to conduct user-based studies on college campuses due to the ease of recruitment, this demographic is not entirely representative of a general population that can utilize multi-factor authentication (Griffin 2015). We found several inconsistencies while recording of the age-group of the participant pool. 68.4% studies provided some variety of formatting for age (E.g, average age, a range of ages, and age groups). Rest of studies never stated the age of their participants but noted that that they were college students or working professionals. Gender studies are often difficult, often leading to imbalanced gender distribution. Previous research regarding gender in usability studies points towards evidence that there is a definitive preference among genders in reference to visual design and usability (Djamasbi et al. 2007). We therefore believe that diverse gender samples in usability studies provide a more accurate depiction of MFA usability in future technological implementations. The average number of male participants is 62.7 and that of female participants is 65.3. This data is highly skewed since, only 12.3% of the papers mentioned gender as a prospective area of research in user studies (Katsini et al. 2018) and 5.3% papers included any gender-based analysis (Cain et al. 2018).

   Educational background information is another fundamental attribute in our meta-analysis and more than half (31 out of 57 papers) of the papers fail to mention any demographic information related to education about the participants. The education distribution throughout all of the user studies primarily shows that most participants were at least college educated when performing the study (Gratian et al. 2018), which again generates recruitment biases. Six of the papers included information regarding the education backgrounds of its participants, and 22 of the studies only included subjects that were either in college or were professionals. Some papers even reported that their participants were Computer Science students as well (Mogire et al. 2018; Shnain & Shaheed 2018). There is very little literature on user studies with individuals who have special needs. Of the 57 papers, only three studies mentioned the need to investigate the disability population for further research (Reynolds et al. 2018). Each of these papers mentioned the usefulness of studying the niche population in authentication, but no paper explored this specific population in depth. Only one paper by Almoctar et al. concluded that its findings would positively benefit the disabled population by providing a MFA scheme that utilizes eye tracking software via webcam to achieve account authentication, thereby foregoing the need for a user to make any kind of physical contact with their device (Almoctar et al. 2018). Only eighteen Gender of the 57 papers mentioned about any compensation given to the participants for completing the study, where the primary compensation included either MTurk rewards (e.g. values less than $1.00) (Kankane et al. 2018) or a small monetary reward (Becker et al. 2018).

| | |
|---|---|
| Male (Average) | 62.7 |
| Female (Average) | 65.3 |
| Gender Based Studies | 3 (5.1%) |
| Mentions Gender For Study | 4 (6.8%) |
| Non-Gender Studies | 52 (88.1%) |
| Education | |
| Various | 5 (19.2%) |
| College | 8 (30.8%) |
| College or Professional | 9 (34.6%) |
| Graduate | 2 (7.7%) |
| Computer Science | 2 (7.7%) |
| Expertise Testing | |
| Technical Expertise Tested | 5 (8.5%) |
| Not Reported | 54 (91.5%) |
| Compensation | |
| Paid Study | 17 (28.9%) |
| Not Reported | 42 (71.2%) |

**Table 4:** Distribution of studies which included demographic details of the participants

### 4.2.4. Methods Used

Core to understanding the trends and gaps within MFA usability research is understanding the varying methodologies and subsequent findings each paper reveals. Even for user-based studies, we found that new authentication technologies comprise a large amount of the existing research. Of the 57 studies, 25 were conducted as studies on newly proposed MFA schemes. Of these 25 papers, 16 studies utilize usability feature testing to assess the performance of their proposed MFA. The rest of the nine studies use in-lab experiments as a means to determine their MFA's effectiveness. Overall, most studies report positive results that are primarily based on enhancing usability (Meng & Liu 2018), improved security (Chithra & Sathva 2018), and increased successful login rates (Irfan et al. 2018). Overwhelmingly, these studies utilize experiments as opposed to surveys, where experiments comprise approximately 76% of the studies that involve new MFA schemes. User behavior and risk perception analysis is yet another large field of research within MFA. Twenty-one papers were based on such research, where eight were conducted in-lab, eleven as online surveys, and two as experiments that used a combination of interviews and surveys. Few studies throughout the papers explored existing MFA schemes. Five of these papers used usability feature testing. These studies outline issues within currently existing MFA, such as usability issues in interface and that better passwords/authentication can only happen when benefits are clear and when users are told to do so. In lab experiments comprise the rest of the six studies, where the overall key findings are that users tend to care about their account security but are not as informed or can recall passwords as well.

## 5. Conclusion

Multi-factor authentication improves online data security by implementing multiple factors in addition to single factor sign-on. Usability of such security technologies often comes across as a challenge for security practitioners, researchers, designers, and developers. Through systematic literature review ($N = 623$) we aimed at understanding the current trends of MFA research and studies. We analyzed the gaps in the existing literature for future user studies ($n = 57$) which can align with the risk perception of individuals. Our study reveals that there are identifiable trends in MFA studies that reveals a considerable amount of focus on new authentication technologies but lacks risk perception analysis. Additionally, we noted that cultural and demographic biases in user study designs. Many studies performed usability testing of existing or proposed new MFA (21 out of 57), however, a two of them discuss implementation and adoption of MFA in large scale organizations. Furthermore, the studies overall show recruitment bias to individuals who come from universities (Khan & Chefranov 2018). We provide actionable recommendations to pave future research scope, primarily aiming to include more diverse population for user study evaluations which can be effective for general adoption of MFA.

## Acknowledgments

This research was supported in part by the National Science Foundation under CNS 1565375, Cisco Research Support, and the Comcast Innovation Fund. Any opinions, findings, and conclusions or recommendations expressed in this material are those of the author(s) and do not necessarily reflect the views of the US Government, the National Science Foundation, Cisco, Comcast, nor Indiana University. We would like to acknowledge the assistance of Sabrina Sanchez who helped us in the initial data collection. We would also like to thank Andrew Kim, Joshua Streiff, and Olivia Kenny for providing feedback for writing this paper.